# Network traffic instability in a two-ring system with automated driving and cooperative merging

*Ziyuan Gu, Meead Saberi*

*Abstract*—In this paper, we characterize the effects of turning and merging maneuvers of connected and/or automated vehicles (CAVs or AVs) on network traffic instability using the macroscopic or network fundamental diagram (MFD or NFD). We revisit the two-ring system from a theoretical perspective and develop an integrated modeling framework consisting of different microscopic traffic models of human-driven vehicles (HVs), AVs, and CAVs. Results suggest that network traffic instability due to turning and merging maneuvers is an intrinsic property of road networks. When the turning probability is low, CAVs do not significantly change the NFD bifurcation, but scatter in both the simulated link fundamental diagrams (FDs) and NFDs reduces leading to higher and more stable network flows. When the turning probability is high, non-cooperative AVs worsen network traffic instability – the NFD undergoes bifurcation long before the critical density is reached. Results highlight the important impact of cooperative merging on network traffic stability when AVs are widely deployed in road networks.

## I. Introduction

THE potential benefits brought by connected and/or automated vehicles (CAVs or AVs) such as increased road capacity and more stable vehicular flows are already demonstrated in the literature [1-3]. However, a few questions remain open with respect to the effects of these vehicles on network traffic flow characteristics. In this paper, we use the macroscopic or network fundamental diagram (MFD or NFD) to provide new insights into one such less explored area on network traffic instability resulting from turning and merging maneuvers.

The effects of AVs on network traffic flow characteristics can be understood using traffic flow theory [1, 4]. Given the arguably shorter and more deterministic computer reaction time compared with that of human drivers, a tighter and more stable vehicular headway can be realized and maintained resulting in an increased road capacity. This is somewhat an expected result with strong intuition, as opposed to characterizing network traffic instability.

### A. Microscopic Car-Following Stability

Two types of car following (CF) stability exist: local (or platoon) stability and string stability [5]. The former refers to the capability of a single vehicle or a platoon of a finite number of vehicles in responding to and absorbing a perturbance, such as a sudden brake, from the leader. If the perturbance causes greater speed fluctuations of the follower(s) with time, or at least such fluctuations do not decay, the system is considered to be locally unstable exhibiting platoon instability. Clearly, this stability definition only applies to microscopic traffic models.

In contrast, the ubiquitous traffic waves represent string instability [2, 6]. By definition, traffic flow is string stable if local perturbations always decay when propagating upstream of arbitrarily long platoons. Compared with local stability, string stability is more restrictive – even if a traffic stream is perfectly free of local oscillations, string instability can still arise [5]. String stability considering AVs and CAVs has recently attracted much attention using different CF models including the optimal velocity model (OVM) [7], the intelligent driver model (IDM) [2, 8, 9], and the empirical cooperative adaptive cruise control (CACC) model [10], among others; see the latter reference for a brief summary.

### B. Macroscopic Network Traffic Stability

While microscopic CF stability focuses on a platoon of vehicles, there is another macroscopic stability concept in relation to the bifurcation and multivaluedness of the NFD[1] [24-26]. There has been some recent efforts to explore the effects of AVs and CAVs on link fundamental diagrams (FDs) [7, 9, 10, 27-30], but from a network perspective pertinent studies are limited especially on the effects of turning and merging maneuvers. One recent attempt was made to analyze the automated NFD resulting from cooperative CAVs at smart intersections [31]. However, the studied traffic traveled in two crossing directions rather than merge into one, which is the same as [32].

By exploring a two-ring system and its two-bin idealization which isolates the effects of turning and merging maneuvers, studies [24, 25] have shown using both analytical and numerical approaches that even for a perfectly homogeneous network with spatially uniform demand, symmetric equilibria with equal flows and densities across all links are unstable, provided that the average network density is sufficiently high. The NFD undergoes bifurcation (i.e., division of a system into two branches) and becomes multivalued beyond the critical density,

Ziyuan Gu is with the Research Centre for Integrated Transport Innovation (rCITI), University of New South Wales (UNSW) Sydney, NSW 2052, Australia. (e-mail: ziyuan.gu@ unsw.edu.au).

Meead Saberi is with the Research Centre for Integrated Transport Innovation (rCITI), University of New South Wales (UNSW) Sydney, NSW 2052, Australia. (e-mail: meead.saberi@ unsw.edu.au).

---

[1] The NFD was empirically observed in a field experiment in Yokohama, Japan [11]. Afterwards it has been widely applied to perimeter or gating control [12-17], congestion or parking pricing [18-21], route guidance [22, 23], etc.

and the network tends to gridlock at a lower jam density. Such bifurcation and multivaluedness of the NFD demonstrates the destabilizing effects of turning and merging maneuvers on network flows, which must be distinguished from the stability concept from a control system's perspective [16].

Network traffic instability was also studied in general networks without the effects of turning and merging maneuvers [33, 34]. For a network consisting of $n$ equal-length links with the same triangular FD, the NFD undergoes bifurcation even before the critical density is reached resulting in a sizable multistate area. The cause is no longer the turning and merging maneuvers but rather the spatially heterogeneous distribution of traffic. In theory, a sufficiently large and congested network can gridlock at a wide range of densities ranging from zero to the jam density, and the NFD can bifurcate and become multivalued even for near-zero densities. Even in the presence of AVs and CAVs, one can still envision the persistent network traffic instability due to the spatially heterogeneous distribution of traffic. But the effect is likely reduced as these vehicles are arguably more adaptive and controllable than HVs in response to road congestion.

A question arises on whether and how network traffic instability is affected by the turning and merging maneuvers of AVs and CAVs in a homogeneously loaded network. While cooperative merging is not a new concept [28, 35-39], studies thus far mainly focused on local effects along freeway segments or merging sections without providing network-level insights. Inspired by previous studies [24-26], network traffic instability due to the turning and merging maneuvers of AVs and CAVs can be potentially characterized and understood using the NFD. Such a macroscopic analysis adds to the existing literature on network flow and stability considerations of these vehicles.

### C. Research Scope

In this paper, we apply a macroscopic perspective of network traffic instability to explore the effects of automated driving and cooperative merging. We revisit the two-ring system to provide some theoretical insights using the NFD and develop an integrated modeling framework consisting of different microscopic traffic models including the human driver model (HDM) [40], the IDM [41], and the cooperative IDM (CIDM) [36]. One of the key findings is that compared with HVs, non-cooperative AVs worsen network traffic instability resulting in relatively earlier NFD bifurcation. However, with CAVs that perform cooperative merging, the arisen instability effect is largely eased, although it never disappears.

## II. TWO-RING SYSTEM: A MACROSCOPIC ANALYSIS WITH AND WITHOUT AUTOMATED DRIVING

The perfectly symmetric two-ring system is arguably the simplest system that isolates the effects of turning and merging maneuvers [24, 25]. While vehicles travel counterclockwise on the left ring and clockwise on the right ring in an indefinite manner, each of them has the same fixed probability of turning and switching to the other ring at both diverging points (i.e., a Bernoulli process). Unlike the seminal studies where the two rings were tangentially connected[2], we set a physical length for the link connecting one ring to the other as a more realistic network representation of diverging and merging. Such a setup helps demonstrate that merging has a significant impact on network traffic instability. At both merging points equal priority is set whenever a conflict arises between one vehicle traveling in its current ring and another switching from the other ring [24, 25].

The two-ring system is physically equivalent to a more general two-region network, provided that vehicles are homogeneously distributed in space within both regions. Link traffic dynamics in each ring, or equivalently network traffic dynamics in each region, can be macroscopically characterized under quasi-equilibrium conditions by the link FD, or equivalently the regional NFD, resulting in a two-bin idealization that permits theoretical analysis. Such an idealization suggests that (i) at quasi-equilibrium density of each ring is approximately homogeneous along its length without being significantly disturbed by the non-signalized junctions; and (ii) vehicles switch from one ring to the other and merge instantaneously. These two assumptions facilitate the macroscopic analysis, but they are relaxed in the microscopic model.

### A. Two-Bin Idealization: Stable vs. Unstable Equilibria

Let us assume traffic flows in both rings follow the same triangular FD $q(k)$ relating space-mean flow and density [27]. Considering shorter reaction times via computer control and tighter headways via platooning of AVs [1, 2, 4], we expect an increase in both the road capacity $q_{cr}$ and the jam density $k_j$[3]. This is illustrated in Fig. 1a assuming both HVs and AVs comply with the same free-flow speed. While the shape of the FD with automated driving may vary depending on how AVs are physically designed, the overall pattern (i.e., the red triangle envelops the black one) shall remain unchanged.

Let $q_x$ and $k_x$ denote the space-mean flow and density of each ring satisfying $q_x = q(k_x)$, $x \in \{1,2\}$. Assuming each ring has a length of $l$ and applying Edie's generalized definition of network flow [42, 43], the average density of the two-ring system is $K = \frac{k_1 l + k_2 l}{2l} = \frac{k_1 + k_2}{2}$ and the corresponding average flow is $Q = \frac{d_1 + d_2}{2lT} = \frac{1}{2}\left(\frac{d_1}{lT} + \frac{d_2}{lT}\right) = \frac{q(k_1) + q(k_2)}{2}$, where $d_1$ and $d_2$ are the total distance traveled in each ring during a time interval $T$. Since vehicles make turning decisions at both diverging points following a Bernoulli process with the same probability $p_{turn}$, the turning flows $q_1^{turn}$ and $q_2^{turn}$ can be macroscopically described by $q_x^{turn} = p_{turn} q(k_x)$, $x \in \{1,2\}$. Thus, macroscopic traffic dynamics of the two-bin idealization can be characterized by a system of differential mass conservation equations:

$$\frac{dk_1}{dt} = \frac{p_{turn}}{l}\big(q(k_2) - q(k_1)\big), \tag{1a}$$

$$\frac{dk_2}{dt} = \frac{p_{turn}}{l}\big(q(k_1) - q(k_2)\big). \tag{1b}$$

Note that Eqs. (1a) and (1b) are no longer valid if one of the rings becomes gridlocked, because then the other ring will also become gridlocked as soon as one vehicle intends to switch to the already gridlocked ring and thus congests the current one[4]. Mathematically speaking, we have $\frac{dk_1}{dt} = \frac{dk_2}{dt} = 0$ when either ring is gridlocked.

---

[2] In other words, the two rings interacted only at the point of tangency where vehicles switch instantaneously.

[3] One can easily see the effect by taking the inverse of the time headway and the spacing, respectively.

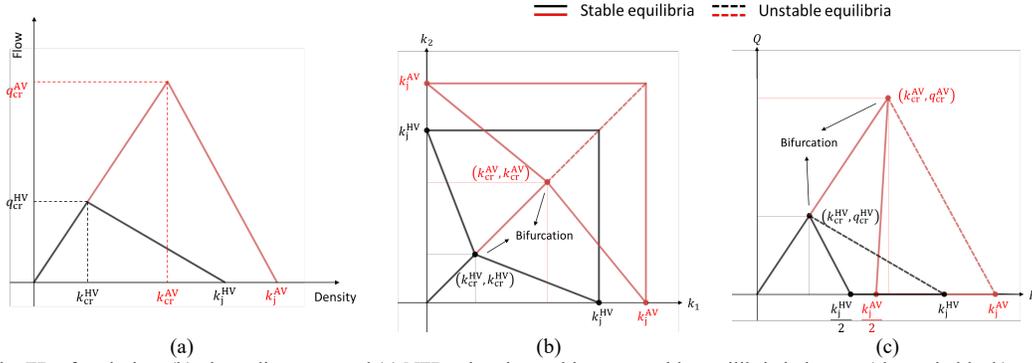

Fig. 1. (a) Triangular FD of each ring, (b) phase diagrams, and (c) NFDs showing stable vs. unstable equilibria in human (shown in black) vs. automated driving (shown in red) environments.

The quasi-equilibrium solutions to the system of equations are straightforward. By setting both derivatives to be zero we have $q(k_1) = q(k_2)$ when $k_1 \neq k_j$ and $k_2 \neq k_j$; otherwise, $k_1 = k_j$ or $k_2 = k_j$. The equation $q(k_1) = q(k_2)$ can be graphically solved by first superimposing both NFDs on the origin but in the opposite directions, and then sliding them to cross each other [24]. While all the resulting intersecting points represent equilibria of the NFD, we must distinguish between stable vs. unstable ones as the latter are self-destructing that cannot sustain themselves. To identify whether an equilibrium point is stable, we observe if any small perturbation from this equilibrium point in $k_1$ or $k_2$ leads to a perturbation of opposite direction in $\frac{dk_1}{dt}$ or $\frac{dk_2}{dt}$.

In Fig. 1b we illustrate all the equilibrium points of the NFD on a phase diagram and compare the results obtained in human vs. automated driving environments. The overall equilibrium pattern remains similar. Before reaching the critical density pair $(k_{cr}^{HV}, k_{cr}^{HV})$ or $(k_{cr}^{AV}, k_{cr}^{AV})$, both rings exhibit unique stable equilibria (see the solid lines) where densities are identical. Beyond the critical density point, however, the equal-density diagonal line becomes the set of unstable equilibria (see the dashed lines) and two branches of stable equilibria form. Thus, we observe bifurcation in the NFD (see Fig. 1c) which leads all the way to a global gridlock at only half of the jam density. By dragging the points $(k_{cr}^{AV}, k_{cr}^{AV})$ and $(k_j^{AV}, k_j^{AV})$ along the diagonal and the points $(k_j^{AV}, 0)$ and $(0, k_j^{AV})$ along the axes, we can visually inspect how the equilibrium pattern shifts on the phase diagram as a result of different design parameters of AVs.

Results suggest that regardless of the frequency of turning and merging maneuvers (i.e., the value of $p_{\text{turn}}$), the resulting network traffic instability persists even in the presence of automated driving. Bifurcation and multivaluedness of the NFD do not disappear and are intrinsic characteristics of network flows. Nevertheless, we may see stability improvement (i.e., delayed bifurcation and less multivaluedness) depending on how AVs are physically designed. As an indicator of how stable the homogeneous two-ring system is against turning and merging maneuvers, we can measure the ratio of the bifurcation density to the jam density. While this ratio is likely to increase in the presence of AVs, we must be aware that the two-bin idealization, as a macroscopic approach, is unable to describe the microscopic dynamics of turning and merging maneuvers. A question thus arises with respect to its effects on network traffic instability. Also, mixed traffic flows cannot be easily characterized as the shape of the NFD varies under different vehicle composition scenarios. Even if the global composition is fixed, allowing vehicles to turn and merge still makes the FD locally variable.

## III. MICROSCOPIC TRAFFIC MODELS: AN INTEGRATED MODELING FRAMEWORK

To explicitly account for the varying longitudinal dynamics of HVs vs. AVs, we model these vehicles separately by their respective CF models. While factors such as safe headway and jam spacing are arguably different between HVs and AVs, the former also differ in several other aspects including reaction time, imperfect estimation, and spatial and temporal anticipation [40]. Further, unlike AVs that are controlled by computers with a deterministic driving style, human drivers are heterogeneous in nature exhibiting stochasticity in their driving behavior. While different CF models such as the OVM [7] and the IDM [2] were applied for HVs (see [10] for a brief summary), we choose the HDM [40] considering its metamodel nature – it can be integrated with a wide range of microscopic CF models[5] – and physical interpretability (see Subsection III.A). Similarly, while there are some other CF models specific to AVs such as those built upon control algorithms [3, 44] and the empirical CACC model [45, 46], we choose the IDM [41] as a widely used model in view of [10] (see Subsection III.B).

Since the two-ring system features a single lane on both rings and both connector links, there is no lane change in the model. However, at both merging points vehicles on the two incoming links need to perform zipper merge into the same outgoing link. This is accomplished by integrating a virtual gap with the CF models [47]. Specifically, when only one of the incoming links are occupied by vehicles, these vehicles will merge into the outgoing link without slowing down or coming to a stop (except that the outgoing link is congested). When vehicles are traveling on both incoming links, the parts of the links within the merging section are virtually superimposed up to the beginning of the outgoing link, so that the leader-follower relation can be identified among any vehicle pairs about to merge. By doing so the virtual gap between the leading and the following vehicles can be obtained, which is then fed into the CF models to calculate the appropriate speed for merging. Based on the calculated virtual gap and following the CF models, vehicles about to merge may need to slow down or even come to a stop before advancing to the outgoing link (e.g., vehicles will stop when the virtual gap is negative).

---

[4] This situation is unlikely to occur in the microscopic model as both connector links have a physical length.

[5] In this paper, the HDM is integrated with the IDM to model HVs.

We assume unconnected AVs are controlled by on-board sensors only and do not communicate between vehicles and with the infrastructure. Thus, they are only able to perform human-like merging maneuvers without cooperation. CAVs, however, are able to leverage the connectivity for cooperative merging. This is accomplished via the cooperative rule of the CIDM [36][6] enforced in the vicinity of both merging points (see Subsection III.C). The focus is on network traffic instability resulting from the merge, which, in fact, can be effectively eased via the cooperative rule (see Section IV).

*A. Human Driver Model*

Building upon the general form of CF models, where the acceleration of the $n^\text{th}$ vehicle $a_n(t)$ is a continuous-time function of its own velocity $v_n(t)$, the net distance $s_n(t)$, and the velocity difference $\Delta v_n(t)$ to its leader, the HDM factors in four critical aspects of human driving including (i) finite reaction times, (ii) imperfect estimation capabilities, (iii) temporal anticipation, and (iv) spatial or multi-vehicle anticipation.

*1) Finite Reaction Times*

As instantaneous reaction time does not apply to human drivers, a finite reaction time $T_\text{react}$ is considered in the HDM. To further account for heterogeneous reaction times of human drivers, we assume a skewed normal distribution with a mean of 1.2 seconds [48].

*2) Imperfect Estimation Capabilities*

Human drivers are subject to perception errors about the surrounding traffic conditions. Such errors affect decision making on acceleration/deceleration and thus contribute in part to traffic oscillations. In other words, the net distance $s_n(t)$ and the velocity difference $\Delta v_n(t)$ are naturally contaminated with noise, whereas the vehicle's own velocity $v_n(t)$ can be read from the odometer with negligible error. Two methods exist to incorporate human perception errors into a CF model [40]. One is by using temporally correlated multiplicative noise and the other by using the conventional white acceleration noise. Although the former introduces a few more parameters with intuitive meanings, the latter is a simpler method that achieves the same effect, which is thus adopted in this paper.

*3) Temporal Anticipation*

Temporal anticipation assumes human drivers are aware of their finite reaction times so that they slightly anticipate the future traffic conditions. To anticipate the future velocity and distance based on the reaction time, a constant-acceleration and a constant-velocity heuristics are applied [40]. The combined effects of finite reaction times, imperfect estimation capabilities, and temporal anticipation lead to the following CF metamodel:

$$a_n(t) = a_n\big(s'_n(t), v'_n(t), \Delta v'_n(t)\big) + \varepsilon(t), \quad (2\text{a})$$

$$s'_n(t) = s_n(t - T_\text{react}) - T_\text{react}\Delta v_n(t - T_\text{react}), \quad (2\text{b})$$

$$v'_n(t) = v_n(t - T_\text{react}) + T_\text{react}a_n(t - T_\text{react}), \quad (2\text{c})$$

$$\Delta v'_n(t) = \Delta v_n(t - T_\text{react}), \quad (2\text{d})$$

where $s'_n(t)$, $v'_n(t)$, and $\Delta v'_n(t)$ are variants of the original variables considering finite reaction times and temporal anticipation, and $\varepsilon(t)$ is zero-mean Gaussian noise.

*4) Spatial or Multi-Vehicle Anticipation*

Assuming human drivers take into account the movements of several vehicles ahead, a CF model can be decomposed into two parts consisting of a single-vehicle acceleration under free-flow conditions $a_n^\text{free}(t)$ and a braking deceleration due to interactions with each of the $k$ preceding vehicles $a_{nm}^\text{int}(t)$ [40]:

$$a_n(t) = a_n^\text{free}(t) + \sum_{m=n-k}^{n-1} a_{nm}^\text{int}\big(s'_{nm}(t), v'_n(t), \Delta v'_{nm}(t)\big), \quad (3\text{a})$$

$$a_n^\text{free}(t) = a\left[1 - \left(\frac{v'_n(t)}{v_0}\right)^4\right], \quad (3\text{b})$$

$$a_{nm}^\text{int}(t) = -a\left[\frac{s^*\big(v'_n(t), \Delta v'_{nm}(t)\big)}{s'_{nm}(t)}\right]^2, \quad (3\text{c})$$

where $a$ is the maximum acceleration, $v_0$ is the desired velocity, and $s^*(t)$ is the desired minimum gap.

*B. Intelligent Driver Model*

The governing equations of the IDM [41] are

$$a_n(t) = a\left[1 - \left(\frac{v_n(t)}{v_0}\right)^4 - \left(\frac{s^*\big(v_n(t), \Delta v_n(t)\big)}{s_n(t)}\right)^2\right], \quad (4\text{a})$$

$$s^*(t) = s_0 + Tv_n(t) + \frac{v_n(t)\Delta v_n(t)}{2\sqrt{ab}}, \quad (4\text{b})$$

where $s_0$ is the minimum distance, $T$ is the safe time headway, and $b$ is the comfortable deceleration. The reason why the IDM is recognized as a suitable model of AVs is threefold [40, 49]: (i) vehicles have instantaneous reaction (depending on the simulation time step size) and perfect estimation of the surrounding traffic conditions; (ii) the term $\frac{v_n(t)\Delta v_n(t)}{2\sqrt{ab}}$ becomes active only in non-stationary traffic streams and enables a collision-free intelligent braking strategy; and (iii) the resulting vehicle dynamics corresponds to natural and smooth driving behavior.

*C. Cooperative Intelligent Driver Model*

To model CAVs that are able to leverage the connectivity for cooperative merging, we introduce the cooperative rule as in the CIDM [36]. It originates from [49] where adaptive cruise control was developed that automatically adapted the driving style to the dynamically changing traffic conditions. Specifically, the change of driving style to facilitate cooperative merging is accomplished by playing with a few multiplication factors on top of Eqs. (4a) and (4b). But such cooperation is only active when a CAV identifies another connected vehicle (either a connected HV or a CAV) within the detection range. In the two-ring system, both merging points have the same detection range (see Fig. 2). CAVs traveling within the detection range automatically adapt their merging behavior whenever another connected vehicle is detected within the range but on the other approach to the merging section.

---
[6] See [28, 35, 37, 38] for optimization- or learning-based merging.

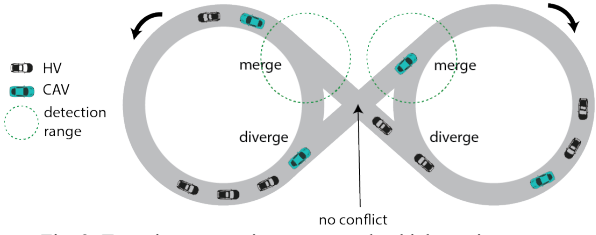

Fig. 2. Two-ring system in a connected vehicle environment.

Four multiplication factors are introduced to help achieve cooperative merging consisting of $\lambda_a$, $\lambda_b$, $\lambda_T$, and $\lambda_s(t)$. These factors are applied one-to-one to $a$, $b$, $T$, and $s_n(t)$ in Eqs. (4a) and (4b). Following [36] we keep acceleration/deceleration of default (i.e., $a$ and $b$ remain unchanged with $\lambda_a = \lambda_b = 1$) while adjusting the headway and distance (i.e., $\lambda_T \neq 1$ and $\lambda_s(t) \neq 1$). Whenever a merging conflict is detected, a CAV immediately changes its driving style by creating a larger spacing from the leader for smoother merging. This is jointly achieved by setting $\lambda_T = 2$ resulting in a larger safe time headway and by applying Eq. (5) to obtain a larger net distance:

$$\lambda_s(t) = \max\left\{0.4, \left(\frac{\Delta x_0(t)}{R_d}\right)^2\right\}, \quad \Delta x_0(t) \leq R_d, \quad (5)$$

where $R_d$ is the detection range and $\Delta x_0(t)$ is the distance to the merging point. Of all the multiplication factors $\lambda_s(t)$ is the only one that varies with time. It gradually decreases as CAVs enter the detection range and approach the merging point until reaching a minimum. After the merge CAVs switch their driving style back to the default (i.e., all the multiplication factors revert to one). The concept underlying Eq. (5) also applies to human drivers who temporally accept tighter spacings as they merge and "relax" shortly afterwards. This relaxation phenomenon was studied elsewhere [50, 51]. Note that CAVs are capable of spatial or multi-vehicle anticipation via the connectivity [36].

IV. SIMULATION EXPERIMENTS, RESULTS, AND DISCUSSIONS

A. Simulation Setup

Following the discussions in Sections I and II we associate AVs with a smaller safe time headway and jam spacing than those of HVs, while the other parameter values remain the same (see TABLE I). The standard deviation of the Gaussian acceleration noise $\varepsilon(t)$ is set to 0.2 m/s$^2$, and the number of spatially anticipated vehicles $k$ is set to three[7].

[7] Calibration and validation of these parameters are not within the scope of this paper.

TABLE I
MODEL PARAMETERS OF HVS AND AVS

| Parameter (unit) | HV | AV |
|---|---|---|
| Desired speed $v_0$ (km/h) | 120 | 120 |
| Safe time headway $T$ (s) | 1.5 | 0.5 |
| Maximum acceleration $a$ (m/s2) | 1.5 | 1.5 |
| Desired deceleration $b$ (m/s2) | 2 | 2 |
| Jam spacing $s_0$ (m) | 2 | 0.5 |

The two-ring system was built using the open source traffic simulator SUMO [52]. The radius of both rings and the length of both connector links are set to 50 m and 100 m, respectively. Vehicles are symmetrically loaded at a constant rate of 180 veh/h onto both rings with a speed limit of 30 km/h. The detection range specific to CAVs at both merging points is set to 30 m. Six replications are performed under each scenario due to simulation stochasticity, each of which lasts for 30 min. The simulation time step as well as the reaction times of AVs and CAVs is 0.1 s. The reaction times of HVs are randomly drawn from the skewed normal distribution.

B. Scenario I: No Turning and Merging Maneuvers

We first restrict vehicles from switching between the two rings to obtain the simulated FDs without the effects of turning and merging maneuvers. By applying Edie's generalized definition of network flow [42, 43] to the simulated trajectory data, macroscopic traffic flow variables are calculated every 10 s for both rings in human vs. automated driving environments (see Fig. 3). The theoretical FDs[8] (see Section II) are obtained using the speed limit as the free-flow speed, the reciprocal of the jam spacing as the jam density, and an estimated congestion wave speed such that the maximal observed flow is enveloped. Alternatively, one can apply the steady-state condition to the CF models [27]. But due to the human factors embedded in the HDM (see Section III.A), such a steady-state analysis fails to produce an accurate result compared with the simulation.

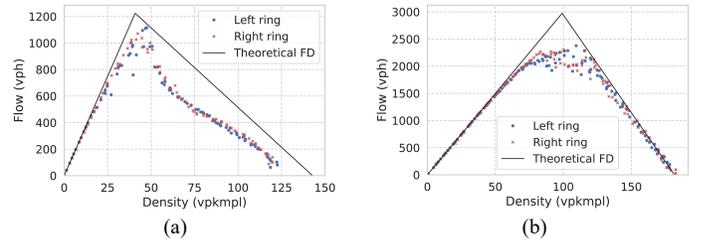

Fig. 3. Simulated FDs without the effects of turning and merging maneuvers in (a) human vs. (b) automated driving environments..

Results suggest that automated driving results in a larger critical density shifting the peak of the FD towards the right. The maximal observed flow or capacity nearly doubles, a not so surprising result consistent with [1]. Note how the simulated FDs associated with HVs deviate from the theoretical one especially in the congested regime immediately after the critical density (where a slight capacity drop occurs). In contrast, the simulated FDs of AVs align quite well with the theoretical one, except for a less notable peak at the critical density. Since turning and merging maneuvers are irrelevant, the revealed discrepancies between the simulated FDs are attributed to the different driving behavior of HVs and AVs – the former exhibit driving heterogeneity and stochasticity whereas the latter feature a homogeneous and deterministic driving style.

[8] The theoretical FDs are not required by the simulation; they only serve for comparison purposes.

### C. Scenario II: Infrequent Turning and Merging Maneuvers

We now consider a low turning probability of 0.15 applied to all the vehicles in the network. Regardless of the vehicle mode, the simulated FD of one ring always evolves towards the congested regime or gridlock, whereas it remains at the free-flow regime for the other ring (see Fig. 4a-c). The resulting NFDs[9] undergo a bifurcation near the critical density and tend to jam at a lower density (see Fig. 4d-f). The result confirms that under congested traffic conditions, simultaneous congested regimes of the two rings are unstable equilibria that cannot self-sustain provided enough time [24]. The network is bound to evolve towards stable equilibria where one ring is congested or gridlocked and the other free flowing.

A more important observation lies in the wide scatter exhibited by the simulated FDs associated with AVs (see Fig. 4b). Unlike those of HVs where scatter only appears around the critical density in a rather small scale (see Fig. 4a), the turning and merging maneuvers of AVs without cooperation result in relatively earlier and greater scatter before the critical density is reached. When comparing the simulated NFDs obtained under both scenarios (see Fig. 4d and e), the simulated bifurcation points (i.e., the onset of the downward trend) are located in the vicinity of the theoretical ones. However, before reaching their respective bifurcation points, the simulated NFDs exhibit different levels of scatter. The upward sloping free-flow branch is nicely reproduced by HVs with almost negligible scatter. But this is not the case for AVs where the simulated points tend to form a local cluster with notable scatter way before the bifurcation point is reached. Thus, lower simulated flows are observed even in the free-flow regime.

By comparing Fig. 4b, c, e, and f, we can immediately observe the stabilizing effects of CAVs on network flows due to cooperative merging. The FD scatter around the critical density reduces leading to a consistently higher and more stable network flows. The maximal observed flow goes slightly beyond 2,000 veh/h across all the replications, which is not the case for non-cooperative AVs as it fluctuates in a wider range that can go as low as 1,500 veh/h.

It is no surprise that the network state represented by the density pair always evolves towards stable equilibria on the phase diagram, regardless of the vehicle mode (see Fig. 5a-c). Before the critical density pair is reached, the simulated phase paths feature almost equally loaded free-flowing rings. However, as soon as congestion builds up in the network, simultaneously congested rings are no longer part of the stable paths and instead, we see spatially asymmetric distribution of vehicles between the two rings. One may notice that the simulated bifurcation points along the phase paths associated with HVs are located beyond the theoretical one (see Fig. 5a). One cause is the difference between the theoretical and simulated FDs (see Fig. 3a) and the other, according to [24], is the faster rate by which vehicles are loaded onto the network that allows insufficient time to ideally reach stable equilibria. Nevertheless, this faster or slower loading of demand does not affect the overall equilibrium pattern.

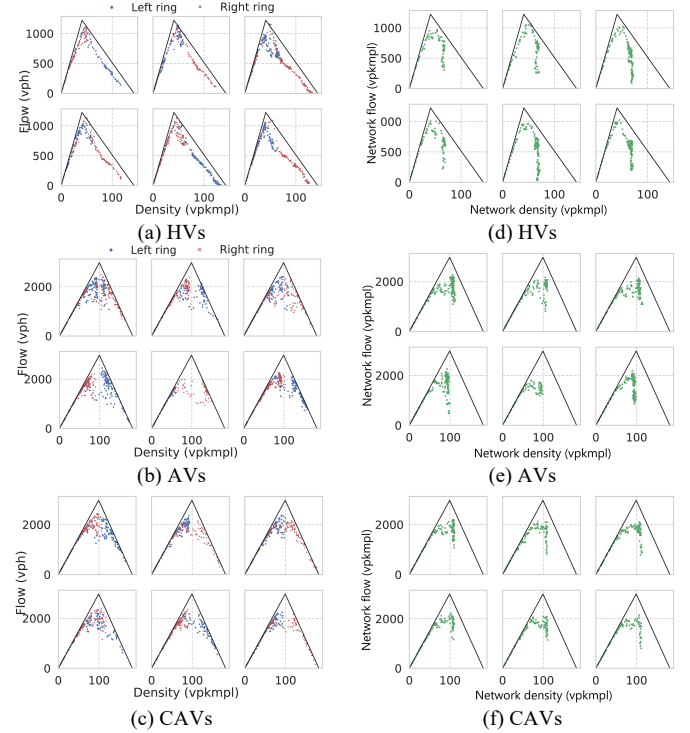

Fig. 4. Simulated (a-c) FDs and (d-f) NFDs of homogeneous traffic flows with 0.15 turning probability.

The simulated phase paths associated with AVs and CAVs are similar since vehicles do not turn and merge frequently (see Fig. 5b and c). But still, some of the paths associated with AVs diverge earlier from the theoretical diagonal and thus, the simulated bifurcation points are closer to the origin. To further quantify the location of a simulated bifurcation point for better comparison across different scenarios, we first smooth the simulated phase path (due to significant fluctuations) via a moving average with a sliding window of 60 s. The bifurcation point is then located if the following two criteria are both met for any two consecutive points $(\bar{k}_1^n, \bar{k}_2^n)$ and $(\bar{k}_1^{n+1}, \bar{k}_2^{n+1})$ along the smoothed path[10]: (i) $|\bar{k}_1^n - \bar{k}_2^n| < |\bar{k}_1^{n+1} - \bar{k}_2^{n+1}|$; and (ii) $\frac{\bar{k}_2^{n+1} - \bar{k}_2^n}{\bar{k}_1^{n+1} - \bar{k}_1^n} < 0$. The former suggests that the smoothed path is diverging from the diagonal, while the latter requires that the slope of the line connecting the two points be negative so as to capture the "bending" characteristic. As shown in Fig. 5d, the simulated bifurcation points associated with AVs and CAVs are closely located in space due to infrequent turning and merging maneuvers. But clearly, they are distant from those of HVs.

---

[9] Connector links are not accounted for to be comparable with the FDs.

[10] $\bar{k}_x$ is the smoothed version of $k_x$, where $x \in \{1,2\}$.

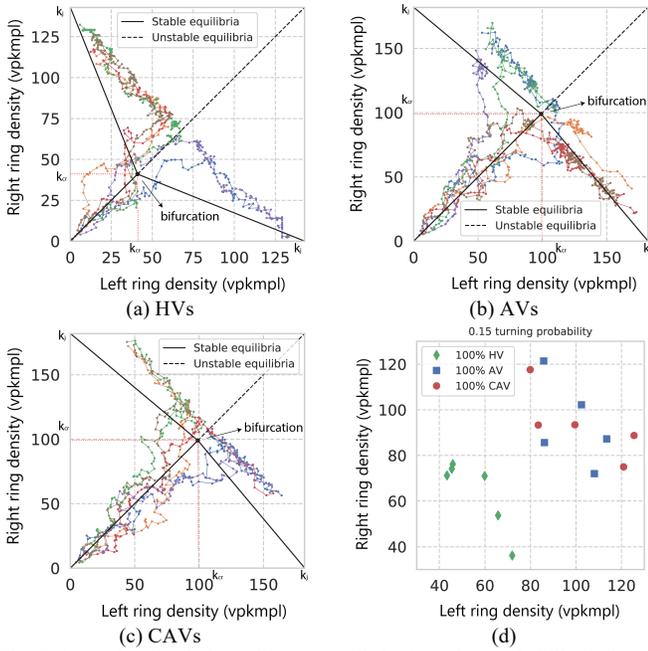

Fig. 5. (a-c) Simulated phase diagrams with 0.15 turning probability in homogeneous traffic flows; and (d) comparison of the simulated bifurcation points.

Overall, when AVs do not turn and merge frequently, leveraging the connectivity for cooperative merging does not significantly change when the network bifurcates in the state space (i.e., the location of the bifurcation point on the phase diagram). However, scatter in both the simulated FDs and NFDs reduces leading to higher and more stable network flows. In the following we will increase the turning probability to further demonstrate the stabilizing effects of cooperative merging.

*D. Scenario III: Frequent Turning and Merging Maneuvers*

We now consider a high turning probability of 0.5 applied to all the vehicles in the network. A comparison between Fig. 6 and Fig. 4 reveals that more frequent turning and merging maneuvers result in greater scatter in the simulated FDs of HVs (see Fig. 6a). Part of this scatter effect arises from the fact that one ring is unloaded and recovers to the free-flow regime. Interestingly, with non-cooperative AVs the transition of one ring from the free flow to the congested regime takes place in a more drastic manner (see Fig. 6b) without exhibiting the same level of scatter as previously observed. The result suggests that one ring undergoes severer performance degradation (i.e., lower network flows) due to frequent non-cooperative merging of AVs that incurs a greater effect of shockwave backpropagation at the merge[11].

The simulated NFDs provide further evidence for the above-described network performance degradation. While the simulated bifurcation points associated with HVs remain in the vicinity of the critical density (see Fig. 6d), those related to AVs move farther away from the critical density and closer to the origin, thus showing an earlier occurrence of the downward sloping congested regime (see Fig. 6e). Accordingly, we observe lower network flows at densities that are far below the critical value. The result once again suggests that network traffic instability is more sensitive to frequent non-cooperative merging of AVs. An alarming drop in network flows is possible even if vehicles in the network are not so many.

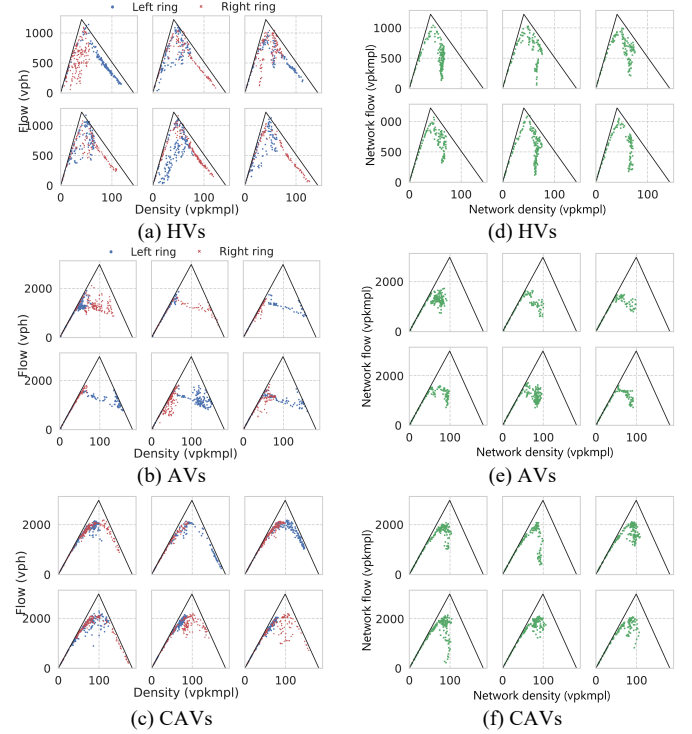

Fig. 6. Simulated (a-c) FDs and (d-f) NFDs of homogeneous traffic flows with 0.5 turning probability.

Now that vehicles are allowed to turn and merge more frequently, the stabilizing effects of cooperative merging via CAVs on network flows become more prominent. One can readily identify such effects by comparing Fig. 6b, c, e, and f, where both the simulated FDs and NFDs become smoother with less scatter. Although the turning probability jumps from 0.15 to 0.5, CAVs are able to maintain the maximal network flow at approximately the same 2,000 veh/h. In contrast, the maximal network flow associated with non-cooperative AVs can go even below 1,000 veh/h (i.e., capacity reduces by half) when densities have yet to reach the critical value. In previous studies on the merging characteristics of CACC [53, 54], we have seen significant capacity drop in the merge bottleneck and thus the necessity of a merging assistance system. Via the cooperative rule embedded in the CIDM (see Subsection III.C), promising results are obtained from the NFD perspective – CAVs can effectively maintain the network capacity and reduce network traffic instability.

We compare the simulated vehicle trajectories along one of the approaches to the merging point. Due to heterogeneous and stochastic human driving behavior, wide and inconsistent spacings between vehicles are observed as well as notable stop-and-go waves (see Fig. 7a). When vehicles are automated, the platooning effect takes place resulting in densified trajectories with small and consistent spacings (see Fig. 7b). However, as we previously discussed, shockwaves persist even in greater scale and propagate backwards at a fast pace. Speed fluctuations do not decay upstream of the platoon showing string instability, which is not the case for CAVs thanks to cooperative merging (see Fig. 7c) – trajectories become smoother without exhibiting stop-and-go waves especially within the detection range.

---
[11] See https://youtu.be/Isz9D440ccI for a video demonstration.

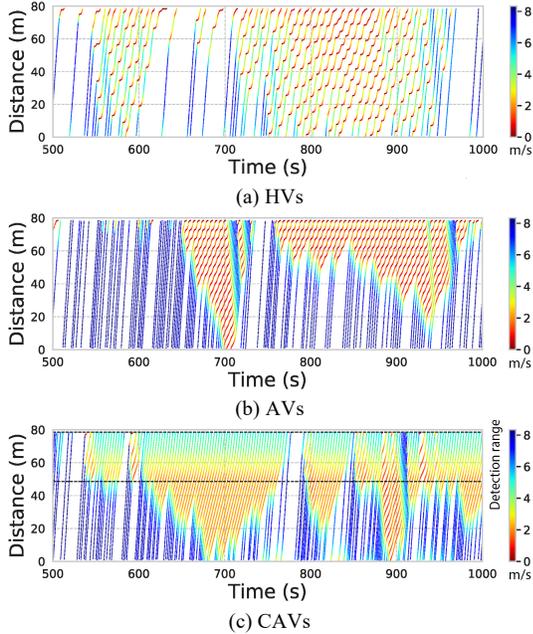

Fig. 7. Simulated vehicle trajectories along one of the approaches to the merging point in homogeneous traffic flows.

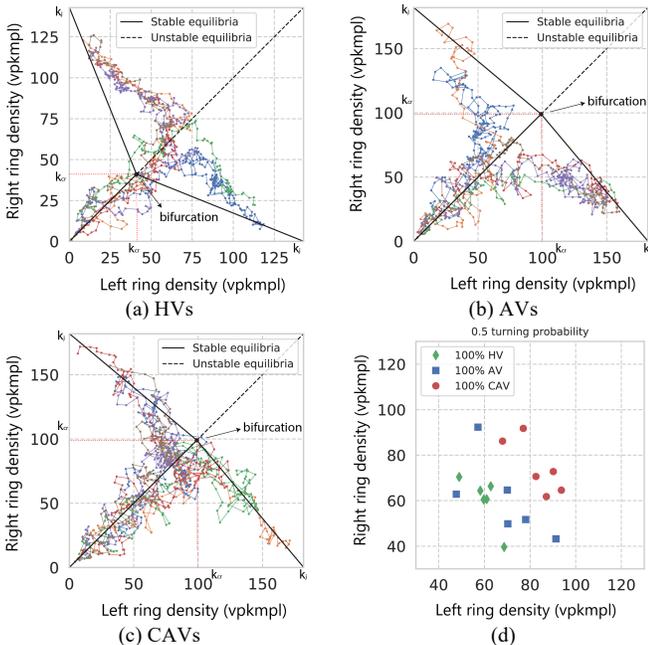

Fig. 8. (a-c) Simulated phase diagrams with 0.5 turning probability in homogeneous traffic flows; and (d) comparison of the simulated bifurcation points.

Due to more frequent turning and merging maneuvers, the simulated phase paths exhibit more fluctuations and are less smooth (see Fig. 8). The paths associated with HVs show a similar pattern to that observed when the turning probability is low (compare Fig. 8a and Fig. 5a). But a drastic change occurs in the paths related to non-cooperative AVs, as they diverge from the diagonal way before reaching the theoretical bifurcation point (compare Fig. 8b and Fig. 5b). The simulated bifurcation points are thus located closer to the origin and, in fact, are in the vicinity of those associated with HVs (see Fig. 8d). In the presence of cooperative merging via CAVs, the paths display a delayed "bending" characteristic and better align with the theoretical stable equilibria (although with fluctuations). Thus, we observe an encouraging overlap of the theoretical and simulated bifurcation points (see Fig. 8c). This explains the visible spatial gap between the simulated bifurcation points pertaining to CAVs and the rest (see Fig. 8d). The farther the points are along the direction between the origin and the jam density pair, the more stable the network flows are under the influence of turning and merging maneuvers.

*1) Heterogeneous Traffic Flows*

Since network traffic instability as well as the stabilizing effects of cooperative merging manifests itself more notably with frequent turning and merging maneuvers, we provide further analysis under this scenario to explore the effects of a varying penetration rate of CAVs (which ranges from zero percent to 100 percent at a step size of 25 percent) in heterogeneous traffic flows comprising either unconnected or connected HVs. The latter can be identified within the detection range so that CAVs can prepare for cooperative merging. But human drivers themselves are assumed to have limited cooperation without changing their default driving behavior. In doing so we are able to focus on the stabilizing effects of CAVs on network flows.

The spatial locations of the simulated bifurcation points on the phase diagram are compared in Fig. 9 under multiple scenarios with a varying vehicle composition. A general trend is observed in the clusters of points, which lie closest to the origin in the absence of CAVs and gradually move farther away as the penetration rate increases. Considering unconnected HVs the spatial differences between the points are not obvious when the penetration rate of CAVs is no greater than 50 percent. Only when it climbs to 75 percent or beyond are we able to identify a visible spatial gap between the points, which suggests that the stabilizing effects of CAVs mixing with unconnected HVs are not effectively realized when the penetration rate is not sufficiently high. It is, however, not rigorous to interpret the result the other way around, concluding that a higher penetration rate of CAVs leads to reduced network traffic instability (although it might be true). The reason is that the ratio of the bifurcation density to the jam density is what constitutes a good stability indicator (see Subsection II.A), where the jam density as the denominator also varies with a changing penetration rate.

When HVs are connected, CAVs with a low penetration rate of 25 percent fail to delay the NFD bifurcation (as in the unconnected case). But the simulated bifurcation points move farther away from the origin when the penetration rate increases to 50 percent and 75 percent. To better demonstrate the effects of unconnected vs. connected HVs, we look separately at each intermediate penetration rate of CAVs (see Fig. 9c) resulting in two observations: (i) when the penetration rate is as low as 25 percent or as high as 75 percent, connectivity of HVs does not make a difference with respect to when the NFD bifurcates; and (ii) when the penetration rate stays at a moderate level of 50 percent, making HVs connected helps delay the NFD bifurcation and thus improves network traffic stability.

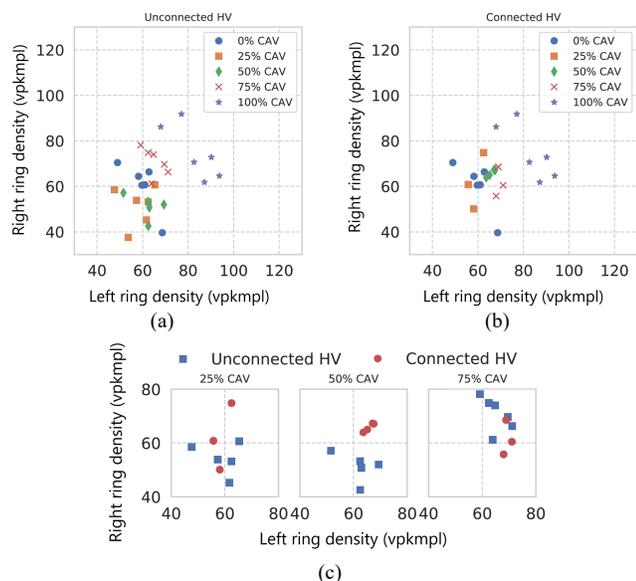

Fig. 9. Effects of a varying penetration rate of CAVs on the spatial locations of the simulated bifurcation points in heterogeneous traffic flows comprising either unconnected or connected HVs.

## V. Conclusion

In this paper, we characterize network traffic instability using a macroscopic approach. Knowing that turning and merging maneuvers of human drivers have a destabilizing effect on network flows, we provide new insights into this effect accounting for automated driving and cooperative merging. This is accomplished by revisiting the two-ring system from a theoretical perspective and developing an integrated microscopic modeling framework for simulation experiments. Results suggest that network traffic instability resulting from turning and merging maneuvers is an intrinsic property of road networks regardless of how (in)frequent such maneuvers are. It does not disappear even if vehicles are connected and/or automated. When the turning probability is low, deploying CAVs provides limited effects on the NFD bifurcation, but scatter in both the simulated FDs and NFDs reduces leading to higher and more stable network flows. When the turning probability is high, the adverse effects of non-cooperative AVs become prominent resulting in worsened network traffic instability (i.e., earlier NFD bifurcation). Under such circumstances, the advantage of CAVs in stabilizing network flows is demonstrated thanks to cooperative merging. Results also suggest that only when the penetration rate of CAVs remains moderate are connected HVs able to provide additional help in delaying the NFD bifurcation.

This study highlights the significance of cooperative merging with respect to network traffic instability, especially when vehicles are automated. The results motivate further thinking on the role CAVs are to play in a complex real-world road network, where merging sections are only part of it known as interrupted facilities. A CAV shall eventually be able to adapt its driving style not only based on nearby traffic flow conditions but also depending on which part of the network it is traveling on. We have seen an increasing trend in recent years towards a reinforcement learning-based approach applied to either uninterrupted facilities [55], merging sections [28, 37, 38], or intersections [56]. But ultimately, an integrated approach is required for automatic and smooth transition between these driving styles.

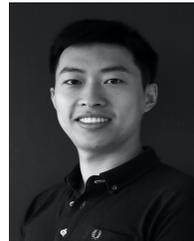

**Ziyuan Gu** received the B.Eng. degree from Nanjing Tech University, China; the M.Eng. degree from SEU-Monash Joint Graduate School, China; and the Ph.D. degree from the University of New South Wales (UNSW) Sydney, Australia.

He is currently a Postdoctoral Fellow with the CityX research lab as part of the Research Centre for Integrated Transport Innovation (rCITI), UNSW Sydney, Australia. His research focuses on transport pricing, network modeling, traffic simulation, and simulation-based optimization.

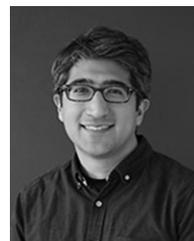

**Meead Saberi** received the B.S. degree from Ferdowsi University of Mashhad, Iran; the M.S. degree from Portland State University, USA; and the Ph.D. degree from Northwestern University, USA.

He is currently a Senior Lecturer in the School of Civil and Environmental Engineering at UNSW Sydney, Australia. He is leading the CityX research lab as part of the rCITI. His research interests cover a range of transportation areas including traffic flow theory and characteristics, large-scale transportation network modeling, complex networks, pedestrian crowd dynamics and simulation, and urban data analytics and visualization.